\documentclass[twocolumn,aps,a4paper,showkeys]{revtex4-1}
\usepackage{graphicx}
\usepackage{url}
\usepackage{float}
\usepackage{fancyhdr}
\usepackage{pslatex}

\begin{document}
\title {Neutron Stars in Teleparallel Gravity}

\author{S. C. Ulhoa}
\email{sc.ulhoa@gmail.com}\email{ulhoa@ualberta.ca}
\affiliation{Faculdade Gama,Universidade de Bras\'{i}lia, 72444-240, Setor Leste (Gama), Bras\'{i}lia, DF, Brazil. \\
Theoretical Physics Institute, Physics Department, University of
Alberta, T6G 2J1,  AB, Canada}

\author{P. M. M. Rocha}
\email{paulomarciano@gmail.com}\email{marciano@ualberta.ca}
\address{Instituto de F\'isica, Universidade de Bras\'ilia
70910-900, Brasilia, DF, Brazil.\\
Theoretical Physics Institute, Physics Department, University of
Alberta, T6G 2J1,  AB, Canada.}

\begin{abstract}
In this paper we deal with neutron stars, which are described by a
perfect fluid model, in the context of the teleparallel equivalent
of general relativity. We use numerical computations (based on the RNS code) to find the
relationship between the angular momentum of the field and the
angular momentum of the source. Such a relation was established for
each stable star reached by the numerical computation once the code
is fed with an equation of state, the central energy density and the
ratio between polar and equatorial radii. We also find a regime
where linear relation between gravitational angular momentum and
moment of inertia (as well as angular velocity of the fluid) is
valid. We give the spatial distribution of the gravitational energy
and show that it has a linear dependence with the squared angular
velocity of the source.
\end{abstract}


\vskip -1.35cm

\maketitle

\thispagestyle{fancy}

\setcounter{page}{1}

\bigskip

\section{Introduction}
\noindent

Most of the known stars are accurately described by Newtonian
physics. Such compact objects as neutron stars with masses of the
order of one solar mass, by contrast, are not. The gravitational field
of such objects is so strong that only general relativity is able to
explain their properties~\cite{1986ApJ...307.178B}. Given that, due to
the very nature of Einstein's equations, analytical methods are very
difficult to implement, most interesting results come from
non-algebraic methods. In particular, to deal with with rotating neutron stars,
one usually resorts to numerical
procedures~\cite{0004-637X-542-1-453,2008IJMPC..19.1863G}.

While the existence of such exotic objects was theoretically predicted
in the 1930's~\cite{PhysRev.55.364,PhysRev.55.374}, the first
observations took place in the
1960's~\cite{1967Natur.216..567P,1968Natur.218..731G,1968Natur.217..709H}.
In neutron stars, the mechanism preventing gravitational collapse is a
repulsive interaction due to the quantum nature of the particles in
the star, which compensates the gravitational pressure, so that
neutron stars exist in very compact spatial volumes. Neutron stars
with masses are around one solar mass and radii are around 10 Km are
known~\cite{RevModPhys.84.25}, as are pulsars rotating with periods as
low as a millisecond and very high surface magnetic
fields~\cite{1992Natur.357..472U,1969ApJ...158....1T}. These very
unusual conditions make them perfect candidates to test any new
predictions of a gravitational theory.

We intend to treat rapidly rotating neutron stars from the
viewpoint of teleparallel gravity.  From the dynamical point of view,
teleparallel gravity and general relativity make identical
predictions. On the other hand, teleparallel gravity allows for the
definition of physically interesting quantities, such as the the
well-behaved gravitational field energy-momentum and angular-momentum
tensors~\cite{Maluf:2002zc}.

In general relativity, the definitions of
these quantities are still under development. The first
attempt to define gravitational energy was based on pseudo-tensors
that were not invariant under coordinate transformations. This effort was
followed by the Komar integrals~\cite{Komar} and then by the formalism
known as the ADM formulation~\cite{ADM}, which is based on
the (3+1)-dimensional Hamiltonian formulation of general relativity,
an approach using constraints to define energy, momentum and angular
momentum. The difficulty with this formalism is that such quantities
are only well-defined asymptotically,  at the spatial infinity. In addition,
no expression obtained in the realm of general relativity
is dependent on the reference frame\textemdash clearly a
feature that is undesirable for energy, momentum, and angular momentum.

The expressions for the energy-momentum and angular momentum of the
gravitational field, in the context of the Teleparallelism Equivalent
to General Relativity (TEGR), are invariant under transformations of
the coordinates of the three-dimensional spacelike surface. They are
also dependent on the frame of reference, as one would expect. Over
the years, they have been consistently applied to many different
systems~\cite{maluf,Maluf:2008yy,Maluf:2005sr,daRochaNeto:2011ir,Ulhoa:2010wv}.

The frame dependence is an expected condition for any expression due
to the field since in special relativity the energy of a particle for
a stationary observer is $m$, but it is $\gamma m$ for an observer
under a Lorentz boost, where $c=1$ and $\gamma$ is the Lorentz
factor. There is no reason to abandon this feature when dealing with
the gravitational field. Similar considerations apply to the momentum
and angular momentum.

Our goal is to establish a relation between the angular momentum of
the gravitational field, which is only predicted by the teleparallel
theory of gravity, and the angular momentum, the angular velocity of
the source, and the moment of inertia. We hope to identify the
significance of the field quantities when related to matter
quantities. In particular, it is known that in the final stage of the
coalescence between two black holes in a binary system the remnant
black hole will acquire linear
momentum~\cite{Pretorius:2007nq,1538-4357-659-1-L5}. As a matter of
fact, the field has to gain the exact amount needed to cancel it in
order to preserve momentum conservation~\cite{Maluf:2012yx}. This
example highlights the importance of knowing the relation between
field and source. We also intend to describe the behavior of the energy
over the entire spacetime and to relate it to other features of neutron
stars.

This paper is organized as follows. In Sec.~\ref{tel} we briefly
discuss certain ideas in teleparallel gravity. In Sec.~\ref{NS} we
describe the space-time of a neutron star and fix the frame that
will be used to compare features only due to the gravitational
field with others due to the source. In Sec.~\ref{AM} we discuss
the field angular momentum and its relation to certain features of the source. In Sec.~\ref{energy} we
relate the gravitational energy to the angular
velocity of the fluid. Plots of the numerical
results are presented to describe the distribution of
gravitational energy in space. Finally, the last section contains
our concluding remarks.

In our notation, space-time indices $\mu, \nu, ...$ and SO(3,1) indices $a,
b, ...$ run from 0 to 3. Time and space indices are indicated
as $\mu=0,i,\;\;a=(0),(i)$. The metric tensor $g_{\mu\nu}$ raises and
lowers space-time indices, while the Minkowski metric,
$\eta^{ab}=diag(-+++)$, acts on SO(3,1) indices.
The tetrad field is denoted by $e^a\,_\mu$, and the determinant of the tetrad field is represented
by $e=\det(e^a\,_\mu)$.  Unless otherwise stated, we adopt units such that $G=c=1$.

\section{Teleparallelism Equivalent to General Relativity (TEGR)\label{tel}}
\noindent

Teleparallel gravity has been investigated over the years as an
alternative to the theory of general relativity (GR). Such a theory
is entirely equivalent to GR as far as the field equations are
concerned. Both theories are derived from Lagrangians that only differ
by a total divergence, and therefore have the same field
equations. The main advantage of TEGR over GR is allowing us
to construct consistent expressions for energy, momentum, and angular
momentum~\cite{Maluf:2002zc}. We will now present ideas and expressions
developed in the realm of TEGR.

The dynamical variables of TEGR are the tetrad fields defined in
the Weitzenb\"{o}ck space-time, also known as the Cartan space-time.
In the Riemann space-time the dynamics is given by the metric tensor.
The tetrad field and metric tensor are related by the equalities
\begin{eqnarray}
g^{\mu\nu}&=&e^{a\mu}e_{a}\,^{\nu}. \label{1}
\end{eqnarray}

Therefore, for each metric tensor one can define infinitely many tetrads.
It follows that the Weitzenb\"{o}ck geometry is less restrictive than Riemann geometry.
The arbitrariness in the choice of the tetrad field is only apparent, since it can be entirely
determined by one more physical condition, namely the observer's reference frame.

As a matter of fact the Weitzenb\"{o}ck and Riemann geometries are
intrinsically related. Consider first the Cartan
connection~\cite{Cartan},
$\Gamma_{\mu\lambda\nu}=e^{a}\,_{\mu}\partial_{\lambda}e_{a\nu}$,
defined in the Weitzenb\"{o}ck space-time. We can equally well write
it as
\begin{equation}
\Gamma_{\mu \lambda\nu}= {}^0\Gamma_{\mu \lambda\nu}+ K_{\mu
\lambda\nu}\,, \label{2}
\end{equation}
where the ${}^0\Gamma_{\mu \lambda\nu}$ are the Christoffel symbols, and
$K_{\mu \lambda\nu}$ is given by
\begin{eqnarray}
K_{\mu\lambda\nu}&=&\frac{1}{2}(T_{\lambda\mu\nu}+T_{\nu\lambda\mu}+T_{\mu\lambda\nu})\,.\label{3}
\end{eqnarray}

$K_{\mu\lambda\nu}$ is the contortion tensor defined in terms of the
torsion tensor $T_{\mu \lambda\nu}$, which is derived from $\Gamma_{\mu\lambda\nu}$. Indeed we have
\begin{equation}
T^{a}\,_{\lambda\nu}=\partial_{\lambda} e^{a}\,_{\nu}-\partial_{\nu}
e^{a}\,_{\lambda}\,. \label{4}
\end{equation}
The tetrad field transforms Lorentz indices into space-time ones and vice-versa, thus $T_{\mu \lambda\nu}=e_{a\mu}T^{a}\,_{\lambda\nu}$.

The curvature of Cartan space-time vanishes identically, as it follows
from substituting $\Gamma_{\mu \lambda\nu}$ in Eq.~(\ref{2}),
which yields the following identity
\begin{equation}
eR(e)\equiv -e({1\over 4}T^{abc}T_{abc}+{1\over
2}T^{abc}T_{bac}-T^aT_a)+2\partial_\mu(eT^\mu)\,,\label{5}
\end{equation}
where $T^\mu=T^b\,_b\,^\mu$, and $R(e)$ is the Riemannian scalar curvature constructed out of the tetrad field.

The density lagrangian of General Relativity (GR), which leaves
the field equations invariant under coordinate transformations, is
precisely the left hand-side of Eq.~(\ref{5}). Given that
any divergence term in a lagrangian density makes no contribution to the
field equations, we drop out the last term on the right-hand side of
Eq.~(\ref{5}) and define the
Teleparallel Lagrangian density as
\begin{eqnarray}
\mathfrak{L}(e_{a\mu})&=& -\kappa\,e\,({1\over 4}T^{abc}T_{abc}+
{1\over 2} T^{abc}T_{bac} -T^aT_a) -\mathfrak{L}_M\nonumber \\
&\equiv&-\kappa\,e \Sigma^{abc}T_{abc} -\mathfrak{L}_M\;, \label{6}
\end{eqnarray}
where $\kappa=1/(16 \pi)$, $\mathfrak{L}_M$ is the Lagrangian
density of matter fields, and $\Sigma^{abc}$ is defined by
the equality
\begin{equation}
\Sigma^{abc}={1\over 4} (T^{abc}+T^{bac}-T^{cab}) +{1\over 2}(
\eta^{ac}T^b-\eta^{ab}T^c)\;, \label{7}
\end{equation}
with $T^a=e^a\,_\mu T^\mu$.

The equivalence between TEGR and GR is now easily understood. Since the
they have the same lagrangian density, the two theories yield the same
dynamics. This does not mean that they predictions are identical.
For instance, there is no quantity equivalent to the
gravitational energy-momentum vector, which will be defined below, in the realm of
General Relativity. In fact every attempt to define the gravitational
energy in the context of GR, from Einstein's equations, has failed. This
approach has only lead to pseudo-tensors in GR.

The field equations can be obtained from the variational differentiation of the Lagrangian density with
respect to $e^{a \mu}$. After a few algebraic manipulations the
resulting equality reads
\begin{equation}
e_{a\lambda}e_{b\mu}\partial_\nu(e\Sigma^{b\lambda \nu})-
e(\Sigma^{b \nu}\,_aT_{b\nu \mu}- {1\over
4}e_{a\mu}T_{bcd}\Sigma^{bcd}) \;= {1\over {4\kappa}}eT_{a\mu}\,,
\label{8}
\end{equation}
where $T_{a\mu}=e_{a}\,^{\lambda }T_{\mu
  \lambda}=\frac{1}{e}\frac{\delta {\mathcal{L}}_{M}}{\delta e^{a\mu
  }}$ is the energy-momentum tensor of the matter fields.

Explicit calculations show that Eq.~(\ref{8}) is equivalent to the
Einstein equations~\cite{maluf:335}.

The field equations can be rewritten as
\begin{equation}
\partial_\nu(e\Sigma^{a\lambda\nu})={1\over {4\kappa}}
e\, e^a\,_\mu( t^{\lambda \mu} + T^{\lambda \mu})\;, \label{10}
\end{equation}
where $t^{\lambda\mu}$ is defined by the relation
\begin{equation}
t^{\lambda \mu}=\kappa(4\Sigma^{bc\lambda}T_{bc}\,^\mu- g^{\lambda
\mu}\Sigma^{bcd}T_{bcd})\,. \label{11}
\end{equation}

$\Sigma^{a\lambda\nu}$ is a skew-symmetric tensor in the last two
indices. This symmetry leads to the following local conservation law
\begin{equation}
\partial_\lambda(et^{a\lambda}+eT^{a\lambda})=0\,\label{13}
\end{equation}
from which it is possible to write down the continuity equation
$$
{d\over {dt}} \int_V d^3x\,e\,e^a\,_\mu (t^{0\mu} +T^{0\mu})
=-\oint_S dS_j\, \left[e\,e^a\,_\mu (t^{j\mu} +T^{j\mu})\right] \,.
$$

We therefore interpret $t^{\lambda \mu}$ as the energy-momentum tensor
of the gravitational field~\cite{maluf2}.  With this concept in mind
we can define the total energy-momentum vector in a three-dimensional
spatial volume $V$ in a familiar way, as
\begin{equation}
P^a = \int_V d^3x \,e\,e^a\,_\mu(t^{0\mu}+ T^{0\mu})\,. \label{14}
\end{equation}

The above-defined energy-momentum vector
is frame dependent and independent of the choice of
coordinates. In order to analyze the features of
gravitational field it is therefore mandatory to set up the reference
frame first. There are countless tetrad fields that match a given metric
tensor. In other words, the same physical system
can be seen under the optics of as many observers as desired.

In the Hamiltonian formulation of TEGR the constraint equations are
interpreted as energy, momentum and angular momentum equations for
the gravitational field~\cite{Maluf:2006gu}. The 4-angular momentum of
the gravitational field can be shown to have the expression
\begin{equation}
L^{ab}=4k\,\int_V d^3x\; e\,(\Sigma^{a0b}-\Sigma^{b0a})\,.
\label{15}
\end{equation}
The $P^a$ and $L^{ab}$ obey the algebra of the Poincaré
group~\cite{Ulhoa}. The above definition, like Eq.~(\ref{14}), is
coordinate independent and changes with the observer.

Although obtained in a Hamiltonian formulation, Eq.~(\ref{15}) is
readily interpretable.  The angular momentum is, of course, the vector
product between the momentum and coordinate. Since the tetrad fields
are the dynamical variables, it is possible to understand the meaning
of Eq.~(\ref{13}) from the perspective of the lagrangian
formalism. The tetrad fields play the role of coordinates, when
contracted with the total energy-momentum tensor, yields the above
equation.

As explained before, the choice of the tetrad field is not random. It
is in fact intimatelly linked to the frame observer. For a given
metric tensor, an infinity of possible frames exists, each of which
completely characterized by the tetrad field. In order to fix the
kinematical state of the observer in the three-dimensional space, we
have to specify six components of the tetrad field, the other ten
components being related to the metric tensor. To this end, we
consider the acceleration
tensor~\cite{Mashhoon1990147,Mashhoon1990176}
\begin{equation}
\phi_{ab}={1\over 2} \lbrack T_{(0)ab}+T_{a(0)b}-T_{b(0)a}
\rbrack\,, \label{16}
\end{equation}
here written in terms of the torsion tensor.

Given a set of tetrad fields, the
translational acceleration of the frame along a world-line $C$
then follows from $\phi_{(0)}\,^{(i)}$, and the angular velocity, from
$\phi_{(i)(j)}$. Consequently, the acceleration tensor is a suitable
candidate to geometrically describe an observer in space-time. It
does not contain any dynamical features dependent on field
equations and has been tested for teleparallel gravity in many
situations~\cite{Maluf:2007qq,Maluf:2009ey,ANDP:ANDP201000168}.

\section{The Space-Time of a Rotating Neutron Star\label{NS}}
\noindent

The most general form of the metric tensor describing the space-time
generated by a configuration with axial symmetry is represented by
the line element~\cite{onda}
\begin{equation}
ds^2=g_{00}dt^2+g_{11}dr^2+g_{22}d\theta^2+g_{33}d\phi^2
+2g_{03}d\phi\, dt\,,\label{17}
\end{equation}
where {\emph all} metric components depend on $r$ and $\theta$:
$g_{\mu\nu}=g_{\mu\nu}(r,\theta)$. If $g$ denotes the determinant
of the metric tensor, we have that $\sqrt{-g}=\lbrack g_{11}g_{22}
(g_{03}g_{03}-g_{00}g_{33})\rbrack^{1/2}$.

For a rotating neutron star, with arbitrary angular
velocity about the $z$-axis, we have the following components of the
metric~\cite{Dinverno,weinberg}:
\begin{eqnarray}
g_{00}&=& -\exp{(\gamma+\rho)}+r^2\omega^2\sin^2\theta\exp{(\gamma-\rho)}\,,\nonumber\\
g_{03}&=&-\omega r^2\sin^2\theta\exp{(\gamma-\rho)}\,,\nonumber\\
g_{11}&=&\exp{2\alpha}\,,\nonumber\\
g_{22}&=&r^2\exp{2\alpha}\,,\nonumber\\
g_{33}&=&r^2\sin^2\theta\exp{(\gamma-\rho)}\,.\label{18}
\end{eqnarray}
Here $\alpha$, $\gamma$, $\rho$ and $\omega$ are metric potentials,
which are functions of $r$ and $\theta$. The matter inside the neutron
star being described by a perfect fluid, thus the energy-momentum
tensor has the form
\begin{equation}
T^{\mu\nu}=(\epsilon+p)U^{\mu}U^{\nu}+pg^{\mu\nu}\,,\nonumber
\end{equation}
where $\epsilon$, $p$ and $U^\mu$ are the energy density, the pressure
and the fluid four-velocity field, respectively~\cite{Dinverno}. Our
assumption of axial symmetry makes $U^\mu$ proportional to the time
and angular Killing vectors. Therefore, $U^\mu\propto(1,0,0,\Omega)$,
where $\Omega$ is the angular velocity of the fluid, measured at
infinity. In this model, to define the metric components in
Eq.~(\ref{18}), we have to specify an Equation of State
(EOS)~\cite{2010IJMPD..19.1569F} relating the energy density to the
pressure and the angular velocity of the fluid.

In order to analyze an axi-symmetrical spacetime within TEGR, we
choose a stationary observer, who has to satisfy
$\phi_{ab}=0$. Adapted to this frame, the tetrad field reads
\begin{small}
\begin{equation}
e_{a\mu}=\left(
  \begin{array}{cccc}
    -A&0&0&-C \\
    0&\sqrt{g_{11}}
\,\sin\theta \cos\phi& \sqrt{g_{22}} \cos\theta \cos\phi & -D\, r
\sin\theta \sin\phi \\
    0& \sqrt{g_{11}}\, \sin\theta \sin\phi&
\sqrt{g_{22}} \cos\theta \sin\phi &  D\, r \sin\theta \cos\phi \\
    0&
\sqrt{g_{11}}\, \cos\theta & -\sqrt{g_{22}}\sin\theta&0 \\
  \end{array}
\right)\,. \label{36}
\end{equation}
\end{small}
The functions $A, C$ and $D$ are given by the equalities
\begin{eqnarray}
A&=& (-g_{00})^{1/2}\,, \nonumber \\
C&=&-{{g_{03}}\over{(-g_{00})^{1/2}}}\,, \nonumber \\
D &=&\biggl[ {{-\delta} \over {(r^2\sin^2\theta) g_{00}}}
\biggr]^{1/2} \,,
\end{eqnarray}
while $\delta$ is defined by the relation
$\delta=g_{03}g_{03}-g_{00}g_{33}$.

We use a numerical method to solve the Einstein equations producing
the metric tensor. Our purpose is to establish a relation
between the features of the gravitational field, such as its angular
momentum, and some experimentally measurable intrinsic attribute of
the neutron star, such as the moment of inertia. Over the years, a
convenient numerical method has been
developed~\cite{1989MNRAS.237..355K},
improved~\cite{1994ApJ...422..227C,1983bhwd.book.....S} and
modified~\cite{Stergioulas:1994ea,Nozawa:1998ak,PhysRevLett.80.4843,0004-637X-502-2-714,0004-637X-591-2-1129}.
We use the RNS code, available at
\url{http://www.gravity.phys.uwm.edu/rns/}, here modified to deal with
such teleparallel quantities as the energy-momentum vector and
gravitational angular momentum. Our calculations use the
non-dimensional quantities listed in Eqs.~(4)-(13) of
Ref.~\onlinecite{1994ApJ...422..227C} and references therein. Here,
however we use $\sqrt{K}$, with $K={c^4}/{G\varepsilon_c}$, as the
fundamental length scale of the system, instead of $K'^{N/2}$, where $K'$ is the
polytropic constant, and $N$ is related to the adiabatic index, which
would be more suitable to deal with polytropic stars,. This
assumes the matter to have no meridional motion, and the angular
velocity $\Omega$, as seen by an observer at rest at spatial infinity,
to be constant.

The program uses compact coordinates, $\mu$ and $s$, defined
by the equality
\begin{equation}
\mu=\cos{\theta}\,,\qquad r=R_e\left(\frac{s}{1-s}\right)\,,\nonumber
\end{equation}
where $R_e$ is $r$ at the equator of the star. Thus $s=0.5$
represents a point on the equatorial surface of the star, while
$s=1$ represents spatial infinity.

\begin{figure}[!ht]
\includegraphics[width=\linewidth]{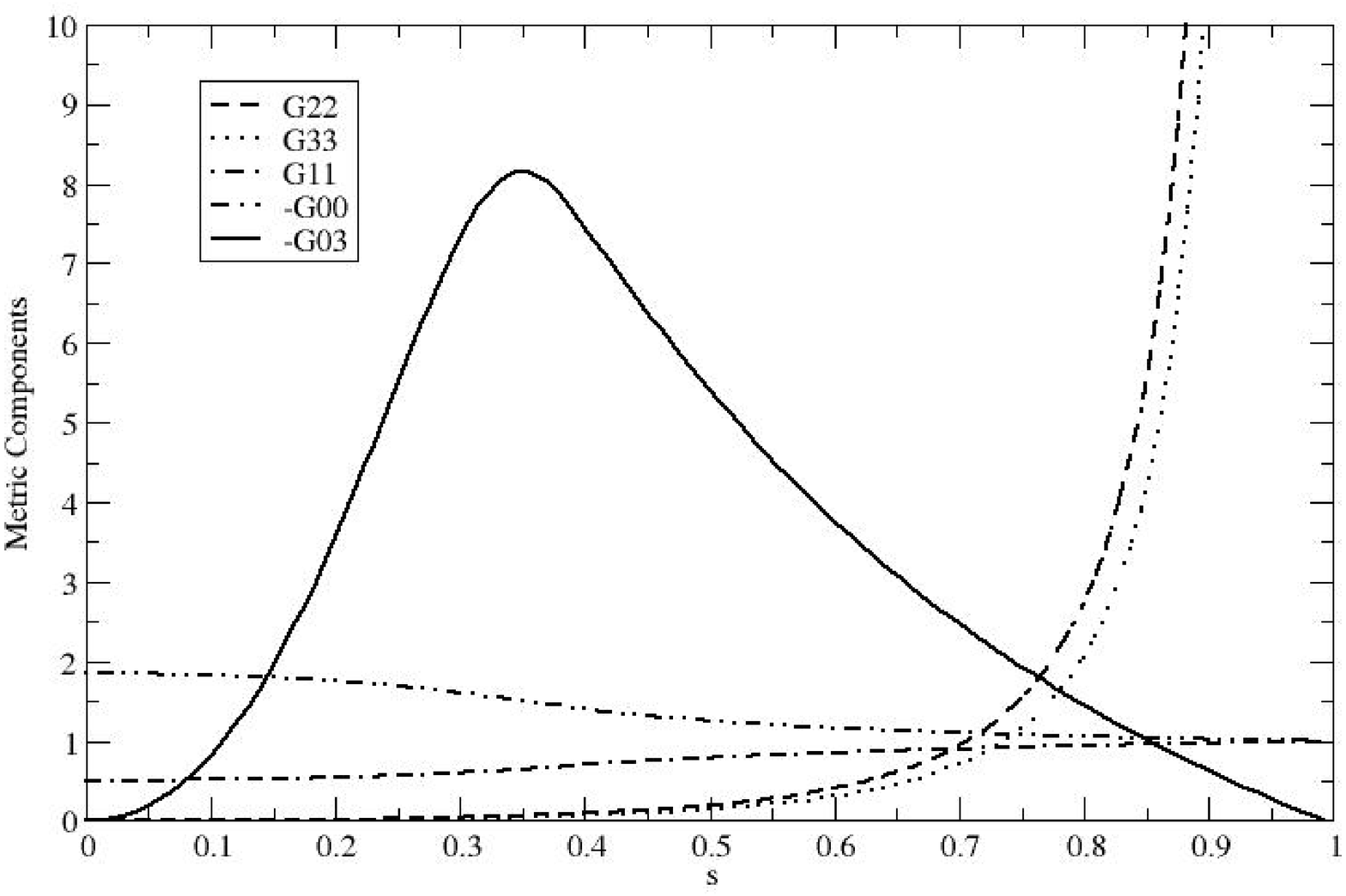}
\caption{ Example of the metric behavior for $\mu=0$, using
EOSA. Here $G_{00}=g_{00}$, $G_{11}=g_{11}$, $G_{22}=g_{22}$,
$G_{33}=g_{33}$ and $G_{03}=g_{03}\,.\, 10^{-3}$.}\label{fig1}
\end{figure}

Figure~\ref{fig1} depicts an illustrative example of the output of the
code. We have chosen the central energy
$\varepsilon_c/c^2=10^{15}g/cm^3$ and adopted
the equation of state in Ref.~\onlinecite{VR1971641}, here denoted by
the acronym EOSA (dense neutron matter). The metric components
indicated in the figure recover the Minkowski
spacetime in spherical coordinates at spatial infinity.

The angular momentum and energy-momentum of the gravitational field
will be discussed in the following sections.

\subsection{The Gravitational Angular Momentum of Neutron Stars\label{AM}}
\noindent

This section uses the results developed in
Ref.~\onlinecite{Maluf:2008ug}. Equation~(40) of that
paper, which relates the $z$-component of the gravitational angular momentum to
the metric, reads
\begin{equation}\label{eq:1}
L^{(1)(2)} = -2k \oint_{S\rightarrow \infty}\,d\theta d\phi \biggl(
{{g_{03}\sqrt{g_{22}}\,\sin\theta}\over \sqrt{-g_{00}}}\biggr)\,.
\end{equation}

Equation~(\ref{eq:1}) describes an axi-symmetrical spacetime, only the
$z$-component of the angular momentum being nonzero, and the analysis
was restricted to slowly, rigidly rotating neutron stars.  Under these conditions,
the angular momentum of the field was found to be proportional to
the (very small) angular momentum of the source. Given that these
conditions are unrealistic, we now want to study rapidly rotating
stars in more complex configurations with a view to determining how
the field angular momentum depends on source features.
We will use the shorthand $L$ for $L^{(1)(2)}$, which will
be dimensionless, unless otherwise stated. To
transform back to the MKS system, all we have to do is
to let $L\rightarrow \frac{c^3 K}{G}L$, where $K$ is the
fundamental length scale used in the numerical simulation.

We now have a computational tool to simulate rapidly rotating neutron
stars and are able to investigate the behavior of the gravitational angular
momentum. We will consider conditions that are more general than those
in Ref.~\onlinecite{Maluf:2008ug}.

\begin{figure}[!htb]
\includegraphics[width=\linewidth]{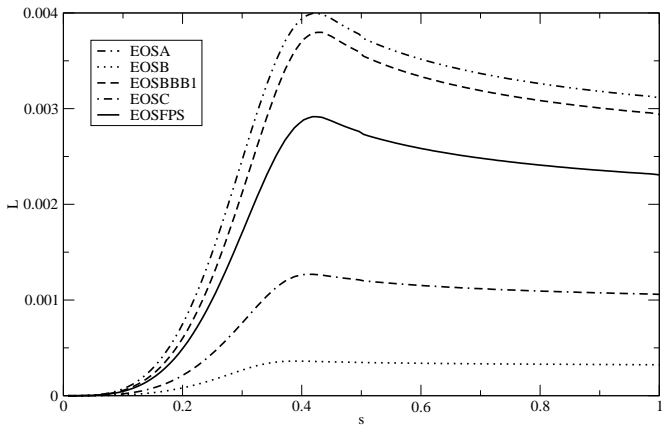}
\caption{Spatial distributions of the gravitational
angular momentum for the indicated equations of state. The
central energy is $\varepsilon_c/c^2=10^{15}g/cm^3$ and the
equations of state are as follows: EOSA~\protect\cite{VR1971641} (dense
neutron matter), EOSB~\protect\cite{1971NuPhA.178..123P} (hyperonic
matter), EOSBBB1~\protect\cite{1997A&A...328..274B} (asymmetric
nuclear matter, derived from the Brueckner-Bethe-Goldstone many-body
theory), EOSC~\protect\cite{Bethe19741} (dense hyperonic matter) and
EOSFPS~\protect\cite{PhysRevLett.70.379} (neutron matter for an
improved nuclear Hamiltonian).}\label{fig2}
\end{figure}

Figure~\ref{fig2} shows $L$ as a function of $s$.  The ratio between
the polar and equatorial radii is $0.6$.  The curves peak
approximately at the surface of the star, then decay smoothly as
$s\to1$, a behavior in agreement with physical intuition, since we
expect $L$ to be maximum near the surface of the star.

\begin{figure}[!htb]
\includegraphics[width=\linewidth]{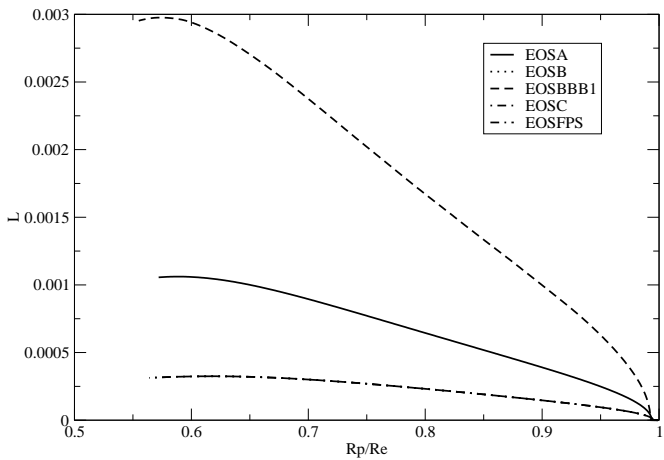}
\caption{Gravitational angular momentum as a
function of the ratio between the polar and equatorial
radii. The legend follows the convention of Fig.~\ref{fig2}}\label{fig3}
\end{figure}

The ratio between the polar radius ($R_p$) and the equatorial radius
($R_e$) defines the shape of the star.  The larger this ratio, the
smaller the gravitational angular momentum should be, since a unitary
ratio, the highest possible value, corresponds to a spherical, static
star. Figure~\ref{fig3} shows $L$ as a function of the ratio $R_p/R_e$.
The plots for EOSB, EOSC and EOSFPS are nearly
coincident. In the intermediate regime,  with $R_p/R_e$ neither too
small nor too close to unity, $L$ is nearly linearly related to $R_p/R_e$.

\begin{figure}[!htb]
\includegraphics[width=\linewidth]{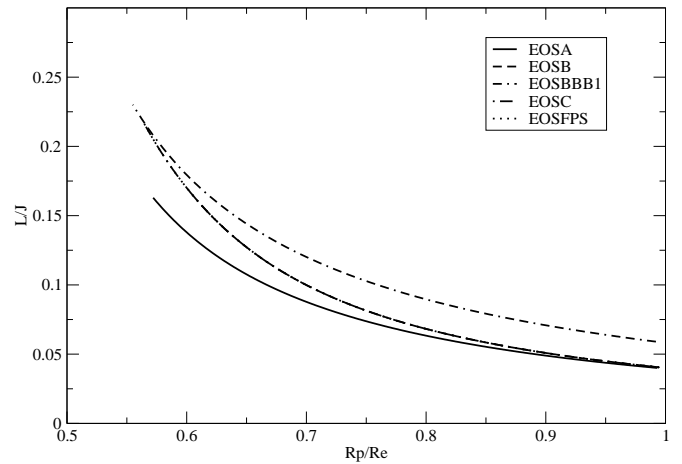}
\caption{Ratio between the gravitational angular momentum and
the angular momentum of the source at the spatial infinity as a
function of the ratio between the polar and equatorial radii. The
legend follows the convention of Fig.~\ref{fig2}.
}\label{fig4}
\end{figure}

\begin{figure}[!htb]
\includegraphics[width=\linewidth]{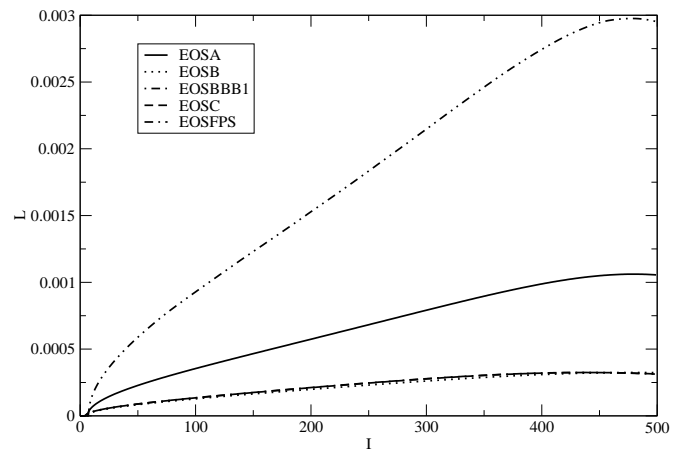}
\caption{Gravitational angular momentum as a function of the moment of
  inertia. The legend follows the convention of
  Fig.~\ref{fig2}}\label{fig5}
\end{figure}

\begin{figure}[!htb]
\includegraphics[width=\linewidth]{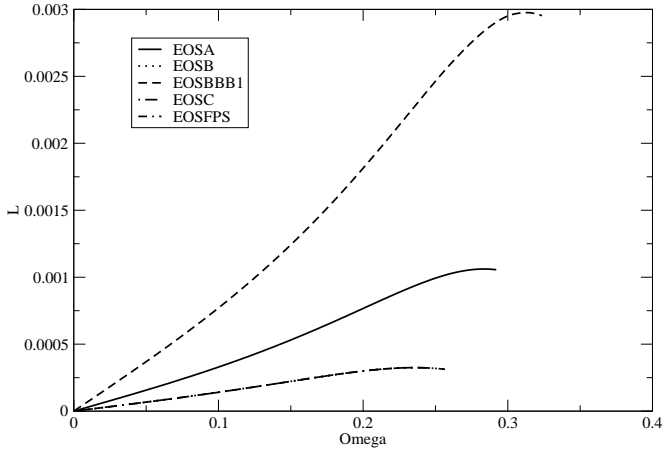}
\caption{Gravitational angular momentum as a function of the angular
velocity of the fluid. The legend follows the convention of
Fig.~\ref{fig2}}\label{fig6}
\end{figure}

We are also interested in comparing the dependences  of the field
($L$) and source ($J$)  on the ratio $R_p/R_e$. We hence plot
$L/J$ as a function of $R_p/R_e$ in Fig.~\ref{fig4}.
Again the curves for EOSB, EOSC and EOSFPS are superimposed,
a feature also found in Figs.~\ref{fig5}~and \ref{fig6}.
Figure~\ref{fig4} shows that in rapidly rotating neutron stars the gravitational
angular momentum is comparable to the angular momentum of the
source. The gravitational field can therefore be detected, since the angular
momentum of the source is experimentally accessible. We
have found that the ratio $L/J$ is proportional to the inverse of
$R_p/R_e$. For a given stable star, $L$ is therefore
proportional to ${R_e}/{R_p}J$.

Our computations make no reference to binary systems. Nonetheless, the
conclusion that $L$ and $J$ have comparable magnitudes shows that the
apparent non-conservation of angular momentum in such systems is not
necessarily linked to the intensity of the emitted gravitational
waves. While beyond the scope of this paper, additional investigation
along this line of reasoning seems undoubtedly interesting.

Figures~(\ref{fig5})~and (\ref{fig6}) present numerical results for
the dependence of $L$ on two other parameters of the star,  the moment
of inertia and the angular velocity of the fluid,
respectively. Figure~\ref{fig5} shows that the gravitational angular
momentum grows linearly with the moment of inertia, except at the
edges of the figure. This is what we expect on intuitive grounds. The edges are non-linear
because they depict regions where $I$  is maximally dependent on $\Omega$.

In Fig.~\ref{fig6}, by contrast, the relation between the
gravitational angular momentum and the angular velocity of the fluid
is non-linear. This is also reasonable, since the angular momentum
varies as the angular velocity of the source grows, because the
rotation is not rigid.

\subsection{The Energy-Momentum Vector of Neutron Stars\label{energy}}
\noindent

For completeness, we now present numerical results for the energy and
momentum.  Given that the metric is stationary, it is reasonable to
expect no energy flux. Even if there is an initial flux, the star will
emit gravitational waves until stability is reached, at which point our
method becomes trustworthy. The combination of the tetrad field in
Eq.~(\ref{36}) with Eq.~(\ref{7}) leads, after some algebraic
manipulation, to the result
\begin{eqnarray}
4e\,\Sigma^{(0)01}&=&\frac {1}{\sqrt {(- g_{00}\delta)}\, g_{22}
g_{11}}  \biggl[2\, g_{11}  g_{22}  \delta- \delta \left( {\frac
{\partial g_{22}}{\partial r}}
\right)\cdot\nonumber\\
&\cdot& \sqrt {g_{22}g_{11} }- \left( g_{22} \right)^{3/2}\sqrt {g_{11} }\,g_{03} \left( {\frac
{\partial g_{03}}{\partial r}}  \right)+\nonumber\\
&+&2\,\sqrt {- g_{00} } \left(
g_{22}\right) ^{3/2}\sqrt { \delta
}\, g_{11}\sin \theta +\nonumber\\
&+& g_{00} \left(  g_{22} \right)^{3/2}\sqrt { g_{11}}\left(\frac
{\partial g_{33}}{\partial r} \right)\biggr]\,.
\end{eqnarray}

Next, on the basis of Eq.~(\ref{14}), we examine the energy as a
function of $s$. We work with dimensionless
quantities and let $E$ denote $P^(0)$. To recover the energy from the
dimensionless variable, we let $E\to(c^4\sqrt{K}/G)E$, where
$\sqrt{K}$ is the fundamental length scale.

\begin{widetext}

First, a simple check on the code. Figure~\ref{fig7} shows the total
energy, i.~e., the energy in a hyper-surface of infinite radius, as a
function of the mass. Although straight over a large fraction of the
displayed range, the solid lines in the first five panels bend up- or
downwards at the highest masses. That the curvatures signal the
breakdown of the nonrelativistic equations of state is shown by the
rectilinear plot in the last panel of Fig.~\ref{fig7}, which contrasts
with the other panels because the EOS in
Ref.~\cite{PhysRevC.58.1804} includes relativistic corrections.

\begin{figure}[H]
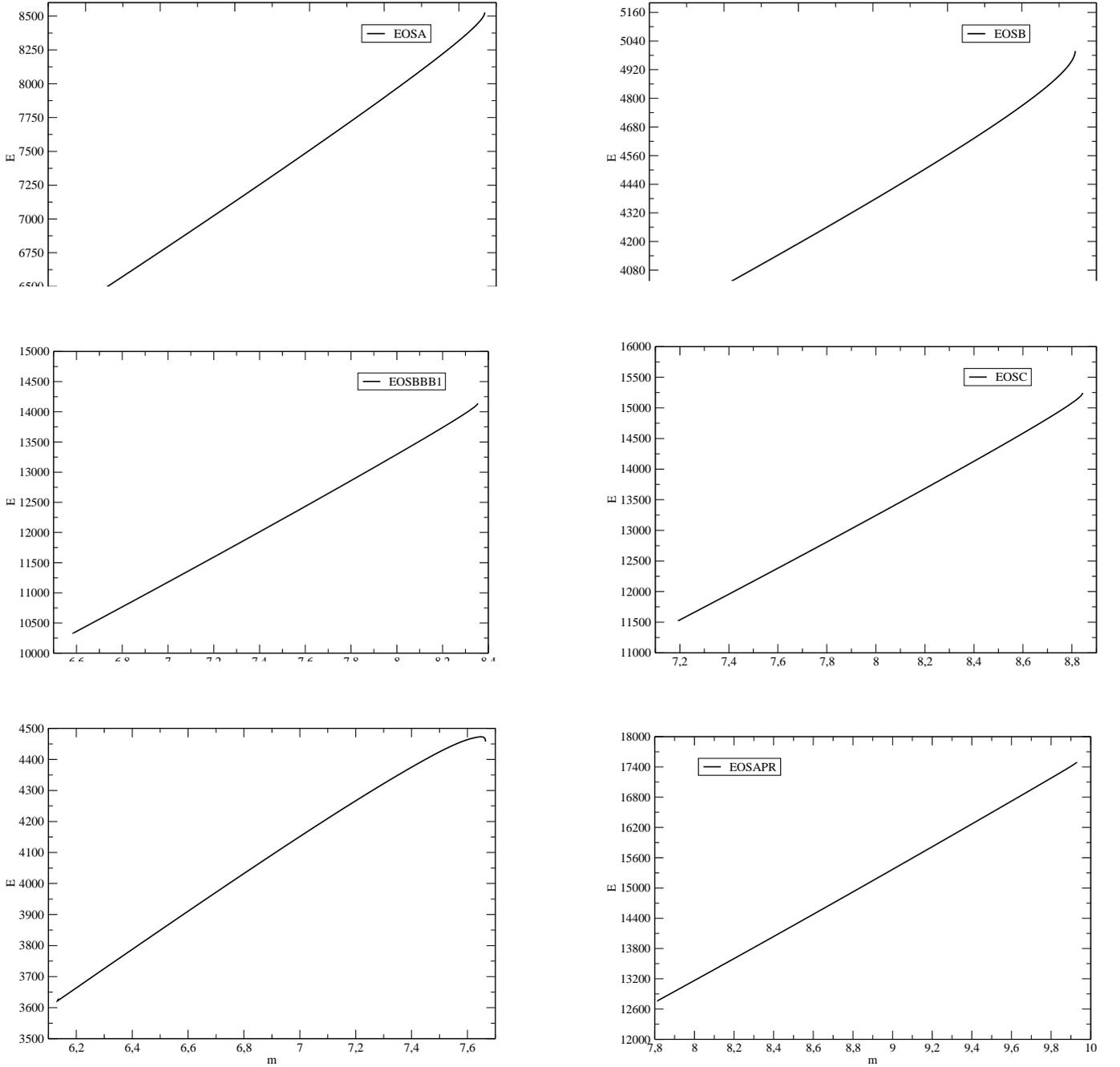

\begin{minipage}[t]{0.45\linewidth}
\includegraphics[width=\linewidth]{EMA.eps}
\end{minipage} \hfill
\begin{minipage}[t]{0.45\linewidth}
\includegraphics[width=\linewidth]{EMB.eps}
\end{minipage}
\begin{minipage}[t]{0.45\linewidth}
\includegraphics[width=\linewidth]{EMBBB1.eps}
\end{minipage} \hfill
\begin{minipage}[t]{0.45\linewidth}
\includegraphics[width=\linewidth]{EMC.eps}
\end{minipage}
\vfill\vspace{0.7cm}
\begin{minipage}[t]{0.45\linewidth}
\includegraphics[width=\linewidth]{EMFPS.eps}
\end{minipage}\hfill
\begin{minipage}[t]{0.45\linewidth}
\includegraphics[width=\linewidth]{EMAPR.eps}
\end{minipage}
\caption{Gravitational energy as a function of the total mass at the
spatial infinity. In this figure $m=10^{11}\,M$, where $M$ is the total mass of the star. The last panel, labeled EOSAPR, has been constructed from the equation of
state in Ref.~\cite{PhysRevC.58.1804}, whci includes relativistic corrections.}\label{fig7}
\end{figure}
\end{widetext}

\begin{figure}[!ht]
\includegraphics[width=\linewidth]{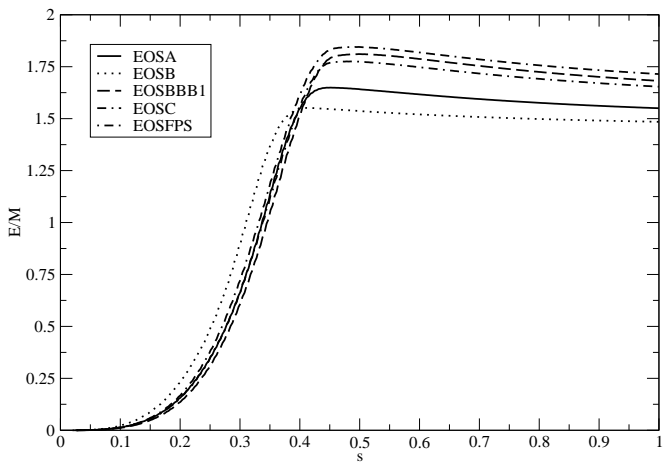}
\caption{Spatial distribution of the ratio between the gravitational
energy and the total mass. The legend follows the convention of
Fig.~\ref{fig2}}\label{fig8}
\end{figure}

Figure~(\ref{fig8}) shows the ratio between the energy and mass as a
function of $s$. For each equation of state, $E/M$ rises up to the
surface of the star and then decreases towards a constant as $s\to1$,
showing that the energy at infinity is proportional to the
mass\textemdash a conclusion that cannot be directly inferred from
Fig.~\ref{fig7}.  The slow decay beyond $s\approx 0.45$ shows that the
rotational energy is very small.

\begin{figure}[!ht]
\includegraphics[width=\linewidth]{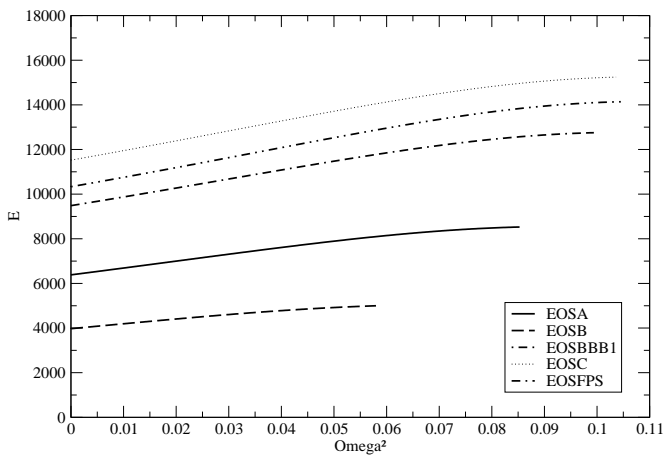}
\caption{Gravitational energy as a function of the
squared angular velocity of the source. The legend follows the convention of
Fig.~\ref{fig2}}\label{fig9}
\end{figure}

Fig.~\ref{fig9} shows the gravitational energy as a function of the
squared angular velocity of the source for each of the indicated
equations of state. As the angular velocity grows, the initially
straight plots bend downards. While, as already explained, the
deviations from linearity in Fig.~\ref{fig7} reflect the inaccuracy of
the non-relativistic equations of state, those in Fig.~\ref{fig9}
have physical origin: they indicate that the moment of inertia, which relates the energy of
the field to the angular velocity, depends on the angular velocity of
the fluid.

We have also computed the momentum of the gravitational field, using the
spatial indices in Eq.~(\ref{14}). The relevant components are
\begin{footnotesize}
\begin{equation}
\frac{4e\,\Sigma^{(1)01}}{\sin \phi}=\left[\frac { g_{03}\left( 2\, g_{22} g_{11} - \left( {\frac {\partial g_{22}}{\partial r}}  \right) \sqrt {
g_{22}\, g_{11}}\right) - \left( g_{22} \right) ^{3/2}\sqrt
{g_{11}}\left(\frac {\partial g_{03}}{\partial r} \right)
 } {\sqrt {- g_{00} } g_{22} g_{11} }\right]\,,
\end{equation}
\end{footnotesize}
\begin{footnotesize}
\begin{equation}
\frac{4e\,\Sigma^{(2)01}}{\cos \phi}=\left[\frac {  g_{03}\left(-2\,  g_{22} g_{11} +\left( {\frac {\partial  g_{22}}{
\partial r}} \right) \sqrt {g_{22}\,g_{11}}\right)+ \left( g_{22}  \right)^{3/2}
\sqrt {g_{11} }\left(\frac {\partial g_{03}}{
\partial r}\right)  }{\sqrt {- g_{00} }g_{22 }
g_{11}}\right]\,,
\end{equation}
\end{footnotesize}
and $\Sigma^{(3)01}=0$.

The momentum is therefore zero, a result that can only be
mathematically understood if we recall that the right-hand side of
Eq.~(\ref{14}) has been integrated over the coordinate
$\phi$. Physically, we expect the rotating star
to transfer no momentum to the field.

\section{Conclusion}
\noindent
We have analyzed the physics of neutron stars from the point of view
of teleparallel gravity. We have determined the energy and angular
momentum as functions of the compact spatial coordinate. In
particular, we have computed the ratio between the field and source
angular momenta. For rapidly rotating neutron stars the magnitudes of
the two angular momenta are comparable. Figure~\ref{fig6} showed that,
as expected, the ratio vanishes in the limit of slow rotations. As
shown by Fig.~\ref{fig6}, at high angular velocities the field angular
momentum ceases to be proportional to the source angular momentum, an
indication that the moment of inertia has become strongly dependent on
the angular velocity of the field.

We have calculated the ratio between the gravitational energy and the
mass as a function of the compact coordinate $s$. This ratio is
maximized near the surface of the star. Beyond the surface, it decays
to a constant at infinity, a value that is weakly dependent on the
equation of state. We are lead to conclude that, as one would expect, each type of neutron
star is characterized by a universal linear relation between the gravitational
energy and mass at spatial infinity. Although we have found small
deviations from linearity for sufficiently large masses,
the comparison in Fig.~\ref{fig7} has linked those deviations to
inaccuracies in the non-relativistic equations of state.

\bigskip
\noindent {\bf Acknowledgement}\par \noindent We thank prof. Sharon
Morsink (University of Alberta) for valuable discussions that led to
the development of this work. We also thank prof. Faqir Khanna
(University of Alberta) for his support and welcome during our visit
at U of A. We thank Theoretical Physics Institute for partial
financial support. One of the authors, P.M.M.R., is also grateful to
CAPES for financial support.


\end{document}